\documentclass[pre,aps,showpacs,nofootinbib,twocolumn]{revtex4}
\usepackage{amssymb}

\usepackage{graphicx}

\input{tcilatex}

\begin{document}

\title{The \ ''unusual '' isotope shift in high-temperature superconductors\\
can be explained by the usual theory of the electron-phonon interaction}
\author{$^{1}$E. G. Maksimov, $^{2}$O. V. Dolgov, and $^{3}$M. L. Kuli\'{c}.}

\address{$^{1}$ P. N. Lebedev Physical Institute,
Leninskii Prosp. 53, 119991 Moscow, Russia, \\
$^{2}$Max-Planck-Institut f\"{u}r Festk\"{o}rperphysik,
Heisenbergstr.1, 70569 Stuttgart, Germany\\
$^{3}$J. W. Goethe-Universität Frankfurt am Main, Theoretische
Physik, Robert-Mayer-Str. 8, 60054 Frankfurt/Main, Germany}

\begin{abstract}
We show that recent ARPES results on the "unusual" oxygen isotope shift in
the real part of the self-energy in the optimally doped $Bi2212$ samples
\cite{nature} can be qualitatively (and semi-quantitatively) explained by
the theory of the electron-phonon interaction (EPI) elaborated few decades
ago. However, for a quantitative analysis of the ARPES spectra it is
necessary to know the momentum dependence of the EPI, the Coulomb
contribution at high energies and the background due to impurities and
defects.
\end{abstract}

\date{\today }
\maketitle

Recently a very interesting paper was reported in the ''Nature'' with the
title \ ''An unusual isotope shift in high-temperature superconductors'' by
Gwon et al \cite{nature}, where the angle-resolved photoemission (ARPES)\
spectra have been investigated in the optimally doped $Bi2212$ samples for
different stages of the oxygen isotope substitution $^{16}O\rightarrow
^{18}O\rightarrow ^{16}O$. It was shown that all energy distribution curves
(EDS) demonstrate a rather small but nonzero isotope effect. Each of these
curves shows a peak, the position of this peak is affected by the above
mentioned isotope substitution. All isotope effects have maximal values for
frequencies of the order of 100-300 meV and vanish for larger energies.

In this short comment we demonstrate that the main part of ARPES\ results in
Ref.\cite{nature} can be easily understood, and even semi-quantitatively
explained, in the framework of the standard Migdal-Eliashberg
model \cite{mig,eli,ES} for the electron-phonon interaction (EPI). For that
purpose we discuss below the quasiparticle self-energy $\Sigma (\omega )$
due to the EPI in the \textit{simple model} in which the quasiparticles interact with a
dispersionless optical phonon (Einstein model), which has been studied at length in Ref. \cite
{ES} many years ago. In this analitically solvable model the Eliashberg spectral function has the form
$\alpha^{2}F(\omega)=\lambda \omega_{0} \delta(\omega-\omega_{0})$
Below we shall discuss the energy and isotope
dependence of $Re\Sigma (\omega )$ for the quasiparticle momenta in the
nodal direction $(0,0)-(\pi ,\pi )$, since in this case there are no
additional effects on $\Sigma (\omega )$ due to the superconducting gap and
the pseudogap. The obtained results in the simple EPI model will be compared with
those of the ARPES spectra in \cite{nature}.

The real part of $\Sigma (\omega )$ in the simple EPI model has the following
form \cite{ES}
\begin{equation}
Re\Sigma (\omega )=-\frac{\lambda \omega _{0}}{2}\ln \left| \frac{\omega
+\omega _{0}}{\omega -\omega _{0}}\right| ,  \label{2}
\end{equation}
where $\lambda =g^{2}N(0/\omega _{0}^{2}$ is the EPI coupling constant, $g$
is the matrix element of the EPI, $N(0)$ is a quasiparticle density of
states on the Fermi level and $\omega _{0}$ is the optical phonon energy.
We would like to mention here that the value $Re\Sigma (\omega )$ is
negative, while $\mid Re\Sigma (\omega )\mid $was plotted in the inset of
Fig.1 in Ref. \cite{nature}.

From Eq.(\ref{2}) it is seen that $\left| Re\Sigma (\omega
)\right| $ in the simple model is logarithmically singular at $\omega \simeq \omega _{0}$. 
It is well known that this
singularity is smoothed by (i) the realistic phonon spectrum and the
Eliashberg spectral function, and (ii) by the nonzero temperature effects $%
T\neq 0$ \cite{Grimvall}. This will be studied elsewhere for spectral dunctions 
related to high-temperature superconductors.

The position of the peak in $\left| Re\Sigma (\omega )\right| $ depends on
isotope substitutions because the phonon energies depend on atomic masses
\begin{equation}
\omega _{0}\varpropto \sqrt{\frac{\gamma }{M}}.  \label{3}
\end{equation}
Here $\gamma $ is a force constant and $M$ is an effective mass of a phonon
mode. The high-energy optical phonon modes in high-$T_{c}$
superconductors are mainly related to the oxygen motion. This means that the
value M in Eq.(\ref{3}) is an oxygen mass. Eq.(\ref{2}-\ref{3}) show the red
shift of the peak position for the heavier isotope. The value of this shift
is equal
\begin{equation}
\Delta \omega =\omega _{0}(^{16}O)-\omega _{0}(^{18}O)\simeq 0.06\omega
_{0}(^{16}O)  \label{4}
\end{equation}
whose order of magnitude is in a good agreement with the observation in Ref.
\cite{nature}.

Now we consider $\left| Re\Sigma (\omega )\right| $ at low energies ($%
\omega \ll \omega _{0}$) where Eq.(\ref{2}) gives a linear behavior
\begin{equation}
\left| Re\Sigma (\omega )\right| \simeq \lambda \omega .  \label{5}
\end{equation}
It is important to stress that in metals $\lambda $ does not depend on the
mass of vibrating atoms in the adiabatic and harmonic approximation, as it
was first shown in \cite{karak}. We stress that this result was observed in ARPES\
spectra of Ref.\cite{nature}, where the slope of $\left| Re\Sigma (\omega
)\right| $ at $\omega \ll \omega _{0}$ is \textit{isotope independent}. It means, 
that the high-temperature superconductors are not in the so called
nonadiabatic regime of the polaron formation. The
expression Eq.(\ref{2}) can be expanded up to the next power of $\omega
/\omega _{0}$ where
\begin{equation}
\left| Re\Sigma (\omega )\right| =\lambda \omega (1-\frac{\omega }{\omega
_{0}}).  \label{6}
\end{equation}
The small deviation (from the linear dependence) of $\left| Re\Sigma (\omega
)\right| $ is negative and is larger for the heavier isotope. This result is
also in a good agreement with the ARPES results shown in the inset of
Fig.(1b) in Ref.\cite{nature}.

Let us now consider $\left| Re\Sigma (\omega )\right| $ at high energies $%
\omega \gg \omega _{0}.$ In this case one has
\begin{equation}
\left| Re\Sigma (\omega )\right| \varpropto \lambda \omega _{0}(\frac{\omega
_{0}}{\omega }).  \label{7}
\end{equation}
This expression shows that $\left| Re\Sigma (\omega )\right| \sim 1/\omega $%
, i.e. it falls off slowly by increasing $\omega $. Moreover \ $\left|
Re\Sigma (\omega )\right| \sim 1/M$, i.e. it depends on the isotope mass. It
means that the value of the isotope effect is more pronounced at high energies then
at low. This result is also in a qualitative agreement with the observed
ARPES data in Ref.\cite{nature}. Note, the same analitical behavior should hold for more 
realistic spectral functions, than for the Einstein model), 
for $\omega \gg \omega _{max}$, where
$\omega _{max}$ is the maximal phonon energy. 

The simple theory presented here cannot explain the absolute value
of the isotope effect at very large $\omega $, which was observed
in Ref.\cite {nature}, since we do not know the contribution of
the Coulomb interaction to $\left| Re\Sigma (\omega )\right| $ at
these energies. However, we can use the difference between the
experimental values of $\left| Re\Sigma (\omega )\right| $ for
$^{16}O$ and $^{18}O$ at rather high energy $\omega =0.2eV$, i.e. $\delta
\left| Re\Sigma (\omega )\right| _{^{16}O}^{^{18}O}\equiv \left|
Re\Sigma (\omega )\right| ^{^{16}O}-\left| Re\Sigma (\omega
)\right| ^{^{18}O}$, in order to estimate the EPI coupling
constant $\lambda _{(h)}$ (in the nodal direction) in the above
simple model. In absence of much more detailed experimental details in this energy region,
we estimate the experimental value of $\delta
\left| Re\Sigma (\omega )\right| |_{^{16}O}^{^{18}O}$ from the small inset in Fig.1
of Ref. \cite{nature} to be
$\delta\left| Re\Sigma (\omega )\right| |_{^{16}O}^{^{18}O}\sim (6-10)$
$meV$, which combined with Eq.(\ref{7}), gives $\lambda _{(h)}\approx
(2-3)$. This value is at least by factor two larger than the coupling
$\lambda _{(l)}\sim 1$
extracted from the slope of $\left| Re\Sigma (\omega )\right| $ at low $%
\omega \ll \omega _{0}$ in the ARPES data 
\cite{nature}, \cite {Lanzara}. Although this  discrepancy in
coupling constants $\lambda _{(l)}$ and$\lambda _{(h)}$ is not too large,
it requires more accurate experiments and
theoretical analysis in the high-energy region.

We stress that for a quantitative analysis of the ARPES data in high-temperature superconductors 
it is necessary
to know (1) the EPI spectral function and its momentum dependence, (2) the
contribution from the Coulomb interaction, (3) the background scattering,
etc. For instance, the point (1) is important in order to explain the
so-called non-shift ARPES puzzle, where the quasiparticle kink in the nodal
direction at 70 meV is unshifted in the superconducting state, while the one
near the anti-nodal point (at 40 meV) is shifted. This non-shift puzzle was
explained in \cite{Kul-Dolg} by the existence of the forward scattering peak
in the EPI and impurity scattering.

In conclusion, the theory of the electron-phonon interaction is capable in
explaining the recent ''''unusual'' isotope shift of the real part of the
self-energy of high-temperature superconductors \cite{nature}, without
invoking exotic mechanisms for the quasiparticle interaction. However, in
spite the fact that recent ARPES data \cite{nature}, \cite{Lanzara}, \cite
{Zhou}, \cite{Cuk} favor the EPI as the pairing mechanism in the
high-temperature superconductors for a quantitative analysis the knowledge
of much more microscopic details are needed.

\end{document}